\documentclass{emulateapj}

\begin{document}

\title{SPECTRAL INDEX OF THE DIFFUSE RADIO BACKGROUND MEASURED FROM 100 TO 200 MHz}
\author{
Alan E. E. Rogers\altaffilmark{1} \&
Judd D. Bowman\altaffilmark{2,3}
}

\altaffiltext{1}{Haystack Observatory, Massachusetts Institute of Technology, Westford, MA 01886, USA}
\altaffiltext{2}{California Institute of Technology, Pasadena, CA 91125, USA}
\altaffiltext{3}{Hubble Fellow}

\begin{abstract}

The mean absolute brightness temperature of the diffuse radio background was measured as a function of frequency in a continuous band between 100 and 200 MHz over an effective solid angle of $\sim\pi$~str at high Galactic latitude.  A spectral brightness temperature index of $\beta=2.5\pm{0.1}$ ($\alpha_S=0.5$) was derived from the observations, where the error limits are $3\sigma$ and include estimates of the instrumental systematics.  Zenith drift scans with central declination $\delta=-26.5^{\circ}$ and spanning right ascensions $0 < \alpha < 10$~h yielded little variation in the mean spectral index. The mean absolute brightness temperature at $\nu=150$~MHz was found to reach a minimum of $T=237\pm{10}$~K at $\alpha=2.5$~h.  Combining these measurements with those of \citet{1982A&AS...47....1H} yields a spectral index of $\beta=2.52\pm{0.04}$ between $150<\nu<408$~MHz.

\end{abstract}


\section{INTRODUCTION}

The last few years have seen renewed interest in the low-frequency radio sky.  This is due, in part, to the development of new radio arrays designed to study the cosmological epoch of reionization (EOR) through redshifted 21~cm emission from neutral hydrogen in the intergalactic medium (IGM) between $6<z<15$.  At these target redshifts, the 21~cm line is shifted well into meter wavelengths and its characteristic rest frequency of $\nu=1420$~MHz is reduced to $202>\nu>89$~MHz.  The detection of the reionization signal is anticipated to be extremely challenging since the low-frequency radio sky is dominated by bright synchrotron radiation and other diffuse emission from the Galaxy, as well as by the integrated contribution of extragalactic continuum sources.  Numerous efforts are underway to investigate strategies for mitigating these foreground contaminants in the planned reionization experiments.  Until new observations are begun, however, these studies must rely either on theoretically motivated arguments for the expected foreground contributions or they must extrapolate measurements from other frequencies, partial sky maps, or limited source catalogs.  Even simple all-sky maps of the Galactic synchrotron contribution must be generated with similar techniques \citep{2008arXiv0802.1525D}.  This situation produces significant uncertainties in the results of redshifted 21 cm foreground subtraction modeling.

One of the most basic measurements needed for extrapolating foreground properties from other frequencies is the spectral index of the diffuse Galactic and extragalactic emission.  In this regard, upcoming redshifted 21~cm experiments overlap with efforts to study and model foregrounds in the latest generations of cosmic microwave background (CMB) experiments. As those measurements become increasingly sensitivity and target more difficult signatures in the CMB, such as polarization anisotropies, the need to quantify and subtract the contributions of diffuse synchrotron and free-free emission from the Galaxy also increases.

The spectral index, $\beta$, of the observed sky brightness temperature is typically defined in temperature units as $T\sim\nu^{-\beta}$.  This is related to flux units according to $S\sim\nu^{-\alpha_S}$, where $\alpha_S=\beta-2$. Much of our fundamental knowledge of the spectral index properties of the low-frequency radio sky originates from the 1960s and 1970s, when there was considerable interest in constraining spatial variations in $\beta$ over the frequency range $10<\nu<1400$~MHz in order to investigate the physical structure of the Galaxy \citep{1962MNRAS.124..297T, 1966MNRAS.132...79A, 1966MNRAS.133..463P, 1967MNRAS.136..219B, 1974MNRAS.166..355W, 1974MNRAS.166..345S, 1976Ap&SS..44..159S}.  At the time, it was recognized that, along with extragalactic sources, the diffuse Galactic radio emission had structure that was composed of three components, originating from the disk, the spiral arms, and a radio halo, respectively \citep{1976Ap&SS..44..159S}. Several experiments were performed using sets of identically scaled horns, antennas, or dipole arrays, in addition to receivers on satellites, to measure accurately the spectrum of the Galactic non-thermal radiation.  The individual components were identified and separated by utilizing analysis tools such as temperature-temperature (``T-T'') plots so that their spectral properties could be investigated, both as functions of frequency and Galactic coordinate. These early measurements resulted in several findings, including that the total diffuse spectrum flattens below 10 MHz \citep{1976Ap&SS..44..159S}, the halo contribution is extremely faint, the spectral index of the disk contribution is dependent on location, and the typical spectral index of the disk contribution steepens rapidly with increasing frequency between 200 and 400~MHz from about $\beta=2.4$ to $\beta=2.8$ \citep{1967MNRAS.136..219B}. More recent investigations \citep{1987MNRAS.225..307L, 1988A&AS...74....7R, 1988A&A...196..211R, 1999A&AS..137....7R, 1998ApJ...505..473P, 2001MNRAS.327..545J, 2003A&A...410..847P} have confirmed many of these early results and demonstrated that the total spectral index saturates to about $\beta=2.9$ above 1~GHz. Recently, there has also been renewed interest in absolute sky temperature measurements around 1~GHz in order to investigate predicted deviations in the CMB from a purely black-body (Planckian) frequency distribution.

\begin{figure*}
\center
\includegraphics[trim=0pc 0pc 0pc 5pc clip]{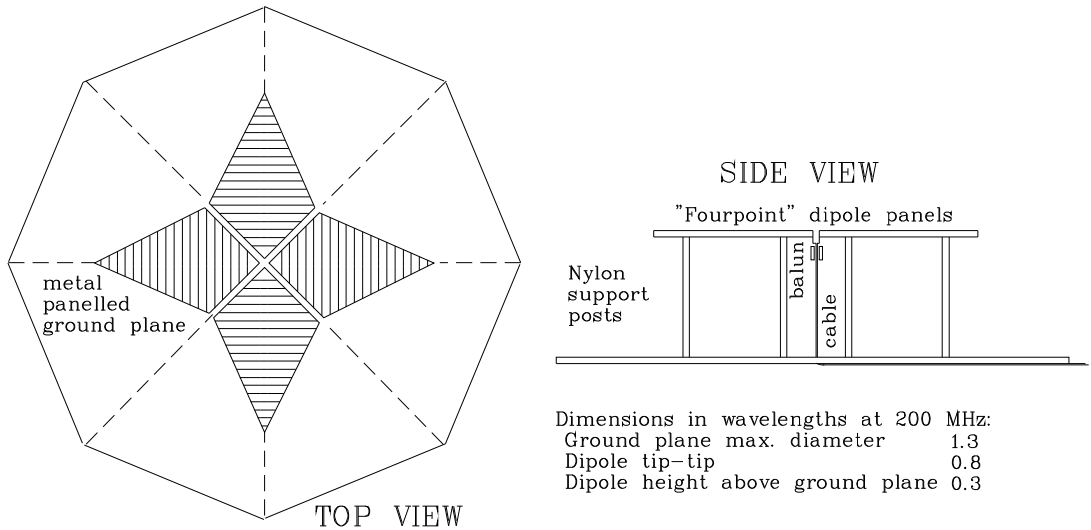}
\caption[Fourpoint antenna]{ \label{f_eorantfig}
Schematic of the EDGES broadband ``fourpoint'' antenna, which is based on the design of \citep{fourpoint2}.}
\end{figure*}

Unlike the early experiments, the modern approach to low-frequency radio instrumentation uses broadband antennas and receivers, along with analog-to-digital sampling systems that are capable of directly sampling radio-frequency waveforms without mixing to intermediate frequencies. We have developed such a system \citep{2008ApJ...676....1B} to cover 100 to 200 MHz for an ``Experiment to Detect the Global EOR Signature'' (EDGES).  This experiment is motivated by theoretical models that predict a weak ($\lesssim35$~mK), but potentially observable, spectral signature similar to a step function due to the redshifted 21~cm contribution to the mean sky spectrum disappearing as reionization unfolds.  Although the design of the EDGES system is optimized for constraining the smoothness of the low-frequency radio spectrum, a small change to the configuration enables the additional calibration needed to constrain the absolute sky temperature and the spectral index of the diffuse emission.

In this paper, we describe the method of absolute calibration and report the results of the absolute temperature and spectral index measurements.  Because the EDGES antenna is a single dipole with a large field of view, it is incapable of performing the difference measurements employed in the pioneering efforts to isolate the Galactic and extragalactic contributions to the spectrum and, therefore, constrains only the total spectrum.  We begin in $\S2$ by describing the experimental approach of the EDGES system.  In $\S3$, we present the procedure for calibrating the sky temperature measurements, along with the values derived from an observing campaign.  We conclude in $\S4$ and $\S5$ with a calculation of the spectral index from the measurements and a summary of previous measurements.

\section{INSTRUMENT DESCRIPTION}

\begin{figure*}

\center
\includegraphics[trim=0pc 5pc 0pc 0pc clip]{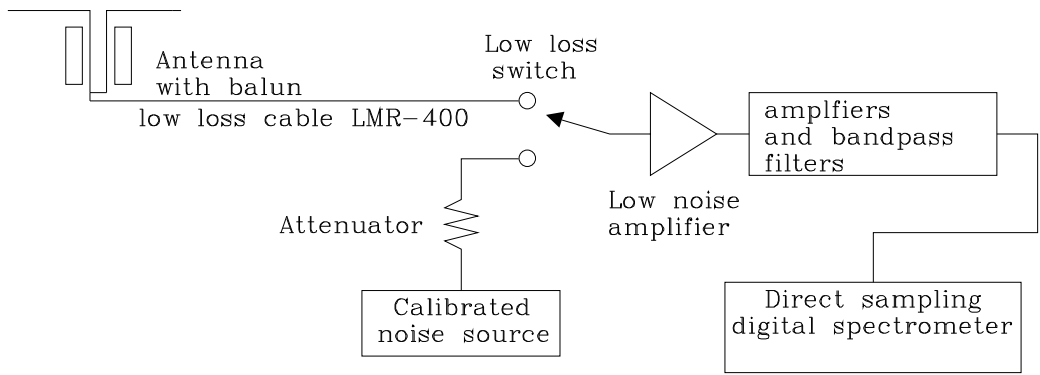}

\caption[Radiometer block diagram]{ \label{f_specindfig1}
Block diagram of the radiometric spectrometer. Spectra are in
a 3-position switching mode which cycles through antenna, load and
load plus calibrated noise.}
\end{figure*}

Previous spectral index measurements used data from a few relatively narrowband observations at widely separated frequencies. A narrowband system can be calibrated in two stages by assuming its properties do not depend on frequency within the observing band.  First, the radiometer is calibrated using a matched load whose temperature is changed from ambient down to that of liquid nitrogen and, second, the radiometer is connected to an antenna with the same impedance as the load. Beyond this it is only necessary to estimate any losses in the antenna. Accurate spectral index measurements can then be made using scaled versions of the antenna---which will have the same impedance and beam pattern, but over different frequency bands.

Accurate absolute measurements with a broadband antenna are more complex because the impedance and beam pattern are both functions of frequency within the observing band. However, over a 2:1 frequency range it is possible to construct an antenna that only deviates a small amount from a perfect impedance match (50 ohm) and whose beam pattern is almost constant over the same frequency range.  For this project, we adapted an antenna, known as a ``fourpoint'', based on the design of \citet{fourpoint2} that exhibits these properties. A schematic view of the antenna as implemented for EDGES is shown in Figure~\ref{f_eorantfig}.  With only small deviations from the ideal narrowband case, it becomes feasible to extend the simple calibration scheme of a narrowband system to this broadband case.

The impact on absolute system calibration of even a small impedance mismatch between the antenna and radiometer is large unless an isolator is placed between the antenna and the receiver to eliminate the effects of correlated noise from the low noise amplifier (LNA) reflected back from the antenna mismatch. Unfortunately, it is hard to make an isolator to cover a 2:1 frequency range so we employed a different strategy.  We used a long, low-loss cable between the LNA and the antenna. The long cable provides sufficient delay in the reflected noise so that the contribution of correlated noise reflected from the antenna can be eliminated in the measured spectrum by averaging over a full cycle of the resulting sinusoidal ripple produced by the beating of the reflected noise with the output noise of the LNA. The amplitude of the correlated noise ripple is proportional to the magnitude of the reflection coefficient, $\Gamma$, whereas the remaining effect of the antenna mismatch after averaging out the ripple depends on $|\Gamma|^2$, which is a much smaller for antennas with a small mismatch ($|\Gamma|\ll1$).  The reflection coefficient of the antenna can be measured with a network analyzer.  Alternatively, the system configuration can be modified by moving a few connections to inject a strong noise source into the third port of a passive 3-port resistive power divider added just before the LNA. In this set-up, the magnitude of the sinusoidal ripple in the spectrum can be used to measure the magnitude of the antenna reflection coefficient as a function of frequency when the antenna is deployed for the sky noise measurements. ``In situ'' measurements are useful as the antenna impedance is influenced, to a small extent, by the ground characteristics beyond the ground plane.

\section{ABSOLUTE SKY TEMPERATURE MEASUREMENT}

\begin{figure*}

\center
\includegraphics[trim=0pc 5pc 0pc 0pc clip]{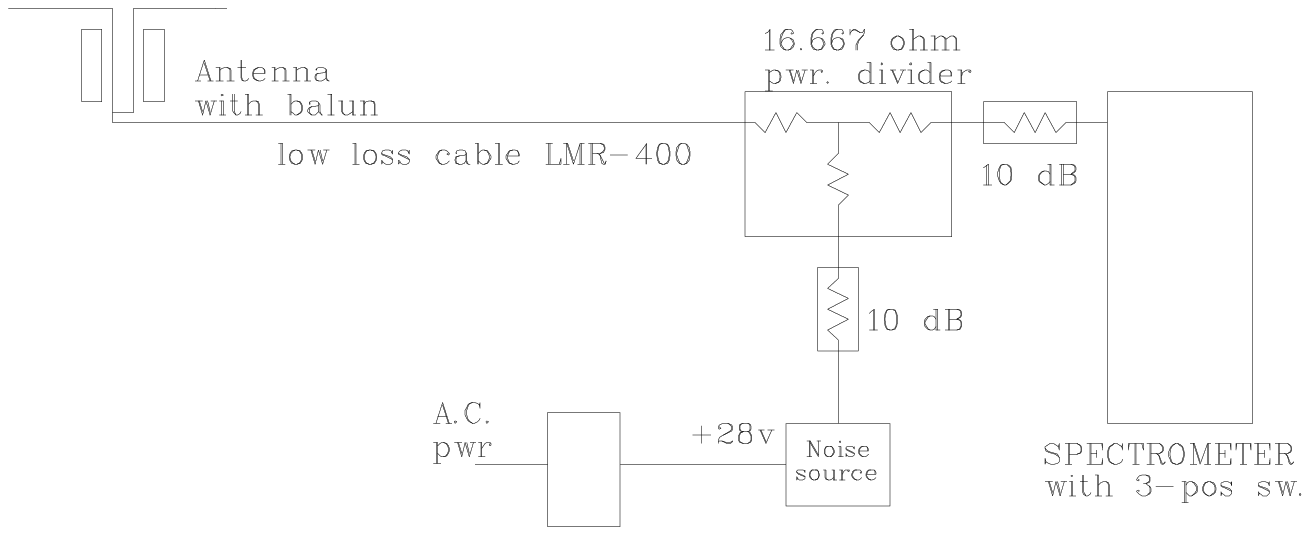}

\caption[Antenna reflection measurement]{ \label{f_specindfig2}
Block diagram of radiometer when used to measure the antenna reflection
coefficient.}
\end{figure*}

Figure~\ref{f_specindfig1} shows the block diagram of the radiometric spectrometer. A low loss coaxial mechanical switch and a calibrated noise source, $T_{cal}$, allows a calibrated spectrum, $T_A$, to be derived from a 3-position switch cycle (ambient load at ambient temperature, $T_L$, load plus calibrated noise) in units of absolute temperature referenced to the antenna input of the switch.
\begin{equation}
T_A = T_{cal}\left[\frac{p_2 - p_0}{p_1 - p_0}\right] + T_L,
\end{equation}
where $p_0$, $p_1$, and $p_2$ are the power spectra on the load only, the load plus calibration noise, and the antenna position respectively. The LNA has a near perfect 50 ohm impedance match and noise temperature of about 50~K.

\subsection{The effect of antenna mismatch}

The effect of the cable and antenna reflection coefficient on the calibrated spectrum from the 3-position switching is given by
\begin{eqnarray}
 T_A = T_{sky}(1 -  |\Gamma(\nu)|^2)L_aL_c + T_L(1 - L_c + L_c|\Gamma(\nu)|^2 - \nonumber \\
 {L_c}^2|\Gamma(\nu)|^2) + T_L(1 - L_a)(1 - |\Gamma(\nu)|^2)L_aL_c +  \nonumber \\
 T_{lna}{L_c}^2|\Gamma(\nu)|^2 + \mbox{sinusoidal terms} ... \; \; \; \; \; \; \; \;
\end{eqnarray}
where $T_{sky}$ is the sky brightness convolved with the antenna beam, $T_L$ is the ambient temperature of the cable and antenna, $T_{lna}$ is the noise out of the LNA input, $L_a$ is antenna loss factor, and $L_c$ is the cable one-way power loss (equivalent to the two-way voltage loss factor).

\subsection{Calibration and Corrections}

The sky temperature measurements were made with amplification module connected
to the antenna via a long (either 15 or 30m) LMR-400
transmission cable. This configuration provides the ability to easily
identify the contribution in the measured spectrum from reflections
due to the impedance mismatch of the antenna and receiver system and,
thus, to calibrate the antenna reflection coefficient. However, it
also requires that the cable transmission coefficient is known since
the ambient load and calibrator noise sources in the amplifier module
are separated from the antenna by the transmission cable.  The losses
due to the ground, balun, antenna, and horizon blockage must also be included
in the analysis and in addition a small correction is needed for the
effect of the noise out of the LNA reflected back from the antenna.
The antenna reflection coefficient and cable
transmission coefficients are dependent on frequency, while the last
set of losses are less dependent on frequency. Below, we
discuss briefly the methods used for acquiring or estimating these
quantities and then present the resulting determination of the
spectral index and absolute temperature of the non-CMB contribution
to the spectrum.

The calibration of the loss due to the transmission cable between the
antenna and the receiver system at the site was determined in much the same way
that the precise temperature of the internal calibration noise source
was measured in the laboratory.  A precision calibrated noise source was connected to
either the amplifier module input directly, or through the
transmission cable. When connected directly, the precision noise
source provides a reference for calibrating the effect of the cable,
as well as a temperature standard for determining the absolute
temperature of the internal noise source. Both the internal noise source
and the precision external noise source had been previously calibrated
in the laboratory using a precision load cooled by liquid nitrogen.
To complete the transmission cable calibration, the transmission cable
was inserted between the precision noise source and the input to the
amplifier module. Measurements of the difference between the spectra
give the loss due to the cable. For the LMR-400 cable, the
power transmission coefficient, $L_c$, was given by the manufacturer
(Times Microwave) to be
\begin{equation}
\label{eqn_edges_transmission}
\log_{10} \left [ L_c(\nu) \right ] = a {\it l}\left(
b~\nu^{1/2}~+~c~\nu \right)
\end{equation}
where $a=-0.003773$, $b= 0.12229$, and $c=2.6\times10^{-4}$, and ${\it l}$ is
the length of the cable measured in meters and $\nu$ is the frequency
in MHz.  For a 30.48 m cable (100~foot), this corresponds to $L_c=0.7184$
and $0.6239$ at $\nu=100$ and $200$~MHz, respectively. Our "in situ" cable loss measurements
were in good agreement with the formula of Eqn. 3. The
internal calibration noise source was found to have frequency-dependent
temperature, $T_{cal}$, according to
\begin{equation}
\label{eqn_edges_tcal} T_{cal}(\nu) = 495 + 30 \left [ \left (
\frac{\nu}{\nu_{150}} \right ) - 1 \right ] K.
\end{equation}
where $\nu_{150}=150$ MHz. This frequency dependence was used in the calculation of
antenna temperature using Eqn. 1.

To determine the antenna reflection coefficient at the observing site, a strong noise
source was injected at the input of the amplification module using a
resistive power splitter as shown in Figure~\ref{f_specindfig2}.
The transmission cable leading to
the antenna remains connected.  This configuration produces
correlated reflections due to the impedance mismatch of the antenna
that are evident in the measured spectrum as a
sinusoidal contribution to the spectrum of the noise source.

The high-level noise source coupled into the coaxial cable to the antenna
produced a spectral ripple amplitude from which the magnitude of
the antenna reflection coefficient was measured.
The calibrated difference spectrum taken between the noise on and noise off
is given by
\begin{equation}
  T_{diff} = T_{noise} \times \left [ 1 + |\Gamma|L_c cos(2\pi\tau+\phi) + \frac{|\Gamma|^2{L_c}^2}{4} \right ]
\end{equation}
where $\Gamma$ is the antenna reflection coefficient, $\phi$ is the reflection phase, $\tau$ is the two-way cable delay, and $T_{noise}$ is the noise source temperature.

\begin{figure}

\includegraphics[width=20pc]{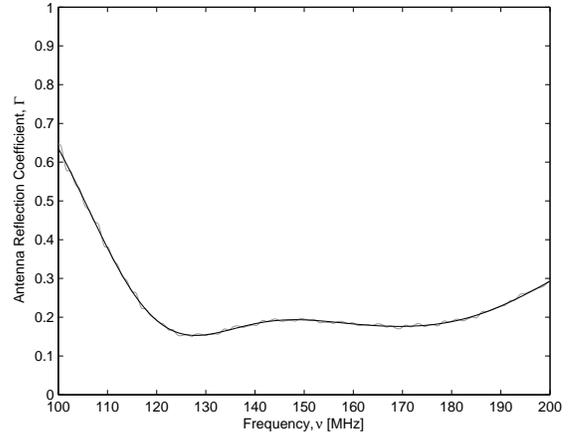}

%\center
%\includegraphics{f4.eps}

\caption[Antenna reflection coefficients]{ \label{f_edges_vswr} Antenna reflection coefficient as a function of frequency.  A $15^{{th}}$-order polynomial is fit (solid) to the raw measurements (gray).}
\end{figure}

To make the reflected noise from the antenna dominant over the sky noise a $10^6$ K noise source was first attenuated by 10~dB to improve its match and then the entire signal attenuated by 10~dB before the 3-position switching LNA input. The reflection coefficient, $\Gamma$, was derived from the ratio, R, of the sinusoidal amplitude to the total amplitude averaged over a period of the sinusoid, according to
\begin{equation}
\Gamma(\nu) = \frac{2}{L_c}\left ( \frac{ 1 - \sqrt{1 -
R(\nu)^2} }{R(\nu)} \right ),
\end{equation}
where
\begin{equation}
R(\nu) = \frac{T_S(\nu)}{T_N(\nu) - T_L} \\
\end{equation}
and $T_S$ is the amplitude of the sinusoidal ripple, $T_N$ is the
measured noise temperature after the contribution of the reflection
has been removed, and $T_L$ is the temperature of the load. In
practice, $T_S$ and $T_N$ are found simultaneously by solving for the
offset and amplitude of a sliding sine wave over a small range of
frequencies corresponding to approximately one period of the ripple.
The period is given by the inverse of the time delay, $\tau_d$, of
the transmission cable.  For the LMR-400 cable,
\begin{equation}
%\tau_d =  \left [ 1 - \left ( \frac{\nu - 200}{500}
%\right ) \right ] \frac{ 2 L }{ 0.85~c},
\tau_d(\nu) =  \frac{ 2 \ell }{ 0.85~c},
\end{equation}
where $\ell$ is the cable length, $c$ is the speed of light, and the
factor of $0.85$ is the relative propagation speed compared to free
space of a wave in the cable\footnotemark.
\footnotetext[1]{Data from Times Microwave Systems. www.timesmicrowave.com/content/pdf/lmr/22-25.pdf}
Thus, the characteristic period in the
measured spectrum of the sinusoidal ripple is
$\tau_d^{-1}\approx4$~MHz for a 30~m cable.
Figure~\ref{f_edges_vswr} shows the antenna reflection coefficient
for EDGES as a function of frequency. The best match (lowest
reflection) is between 130 and 200~MHz.  The reflection coefficient
was also measured using a network analyzer at Haystack Observatory
with similar results.

Lastly, the balun, ground-screen, antenna, and horizon losses are
determined using either laboratory measurements or numerical
simulations. For the simple balun used with EDGES, the loss can be
measured in the laboratory by connecting two identical balun
assemblies back-to-back, with balanced connections reversed in polarity
at the interface, and measuring the transmission. The loss due
to the 572 $\Omega$ ferrite core choke balun to a coaxial cable is
found using this method to be $0.1\pm{0.05}$ dB at the
frequencies of interest. The losses due to the finite size of the
ground screen are more difficult to estimate. Numerical simulations using
the numerical electromagnetics code (NEC) Sommerfeld-Norton high
accuracy ground model were performed to estimate these contributions.
The ground loss was found to be about 0.25~dB for the ground screen at
150 MHz and was found to vary inversely with frequency.
To further constrain the estimated
loss due to the ground screen, measurements were made in the field
after extending the ground screen to a diameter of about 3.5~m using
aluminum foil. The NEC simulations with the extended ground screen
predicted a loss of 0.07 dB. These measurements are discussed in the analysis
below.  The resistive loss in the antenna due to the finite resistance of
the aluminum panels was also modelled and it was found to be
about 0.01~dB. The loss due to the presence of objects blocking the
horizon was estimated from the beam pattern to be about 0.05 dB.

The contribution of all the calibration corrections determines the
total systematic uncertainty in the derived properties of the radio
spectrum.  Table~\ref{t_edges_cal_sum} summarizes the corrections
discussed in this section and the uncertainty in these corrections.
Errors in these corrections result in errors in the spectral index
and the sky noise at 150 MHz derived from these measurements. These
systematic errors dominate the uncertainty in the derived spectral index.

\begin{deluxetable*}{cccl}
    \tablecaption{Calibration Corrections and Uncertainties}
    \tablewidth{0pt}
    \tablehead{
        \colhead{Source} &
        \colhead{Correction} &
        \colhead{Uncertainty} &
        \colhead{Method} \\
        }
    \startdata
    $\Gamma$ & see Fig.~\ref{f_edges_vswr} & $<0.05$ & measured in field \\
    $L_c$ & see Eqn.~\ref{eqn_edges_transmission} & $<0.05~dB$ & measured in laboratory \\
    $T_{cal}$ & see Eqn.~\ref{eqn_edges_tcal} & $<10~K$ & checked with precision source \\
    horizon loss & 0.05 dB & $<0.02~dB$ & model \\
    ground loss at 150 MHz& 0.25 dB & $<0.05~dB$ & NEC Sommerfeld-Norton \\
    extended screen loss & 0.07 dB & $<0.05~dB$ & model \\
    antenna loss & 0.01 dB & $<0.02~dB$ & model aluminum skin resistance \\
    balun loss & 0.10 dB & $<0.05~dB$ & measured in laboratory \\
    noise out of LNA input & 40 K &$<10~K$ & measured in laboratory \\
    \enddata
    \label{t_edges_cal_sum}
%    \tablecomments{ \label{t_edges_cal_sum}
 %   }
\end{deluxetable*}

\section{RESULTS}

Combining the corrections above to calibrate observations performed
with EDGES yields accurate determinations of the absolute sky temperature. To
study the foreground contribution to the spectrum, which we
refer to as $T_{gal}$ to distinguish it from the CMB
contribution, we employ the model
\begin{equation}
T_{gal}(\nu) = T_{150} \left ( \frac{\nu}{\nu_{150}} \right
)^{-\beta}
\end{equation}
where $T_{150}$ is the temperature at $\nu_{150}=150$~MHz, and $\beta$ is the spectral index. For a model of the complete sky temperature, we neglect any redshifted 21~cm contribution and treat the Galactic and extragalactic components of the foreground spectrum together in $T_{gal}$, so that
\begin{equation}
T_{sky}(\nu) = T_{150} \left ( \frac{\nu}{\nu_{150}} \right
)^{-\beta} + T_{cmb}.
\end{equation}
Taking $T_{cmb}=2.725$ \cite{1994ApJ...420..439M}, we solve for $\beta$ and $T_{150}$.

\begin{figure}

\includegraphics[width=20pc]{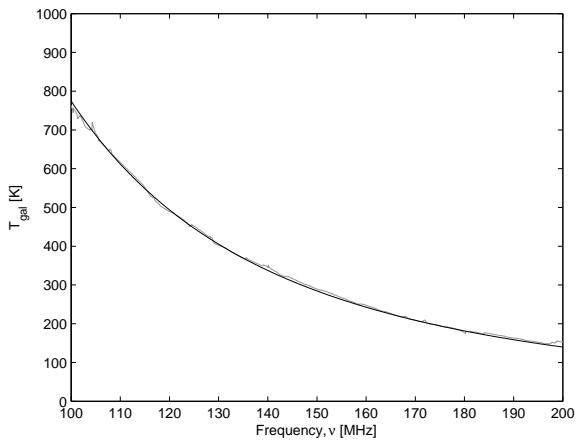}

%\center
%\includegraphics{f5.eps}

\caption[Example calibrated sky spectrum]{ \label{f_edges_gal_spectrum} Example calibrated sky spectrum (gray)
after removal of RFI and subtraction of $T_{cmb}=2.725$ compared to model fit (solid line).  One 25~s integration cycle from observations in the north-south polarization with a 30m transmission cable at approximately 8~h LST on 7 Dec 2006 was used for this plot.}
\end{figure}

Calibrated measurements to accurately determine the spectral index between 100 and 200~MHz were acquired while the EDGES system was deployed at Mileura Station in Western Australia between November 29 and December 8, 2006. Figure~\ref{f_edges_gal_spectrum} illustrates an example fit of the model to a typical observation made on the night of December 7, 2006, with the approximately north-south polarization of the antenna. It is clear in Figure~\ref{f_edges_gal_spectrum} that the model is a good fit to the measurements yielding, in this case, $\beta=2.470$ and $T_{150}=283.20$.

We derive the expected error in the spectral index and absolute temperature estimates using the dependencies listed in Table~\ref{t_edges_parm_sens} on the assumption that the systematic errors add in quadrature. If we take the loss corrections in quadrature the expected error is 0.09~dB and add to this an uncertainty of 3~K in the ambient temperature and an uncertainty 0.05 in the voltage reflection coefficient the combined effect is an uncertainty of $\Delta \beta=0.08$ in spectral index and $\Delta T_{10}=5$~K in the sky noise at 150~MHz. The effect of uncertainty in the noise emitted out of the LNA toward the antenna is negligible.

\begin{figure}

\includegraphics[trim=1pc 6pc 0pc 20pc, clip, width=23pc]{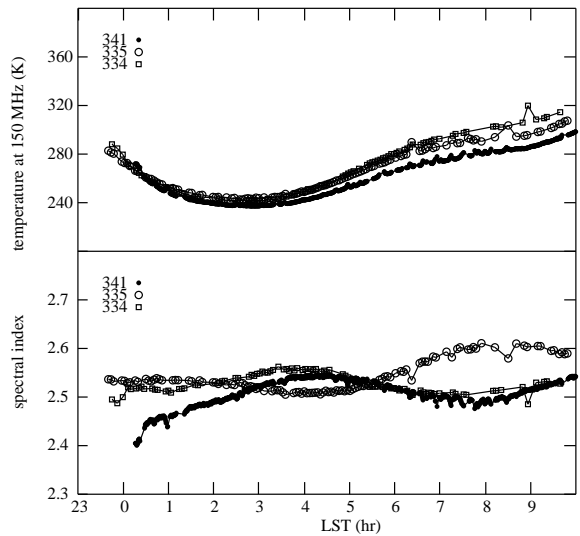}

%\center
%\includegraphics[trim=0.6in 1in 0 3.5in, clip]{f6.eps}

\caption[Sky temperature and spectral index as functions of LST]{ \label{f_edges_gal_results} Derived spectral index between 100 and 200~MHz (bottom panel) and $T_{gal}$ at 150 MHz (top panel) as functions of LST . Three data sets were used in the plots. The data points indicated with small filled circles are for observations on 7 Dec 2006 using a 30m transmission cable and east-west polarizations, while the squares are from 30 Nov 2006 and the open circles are from 1 Dec 2006, both of which used 15~m cables and the north-south polarization of the antenna.  For the 1 Dec 2006 observations, the ground screen was extended to test the contribution of ground loss.}
\end{figure}

\begin{figure}

\includegraphics[width=20pc]{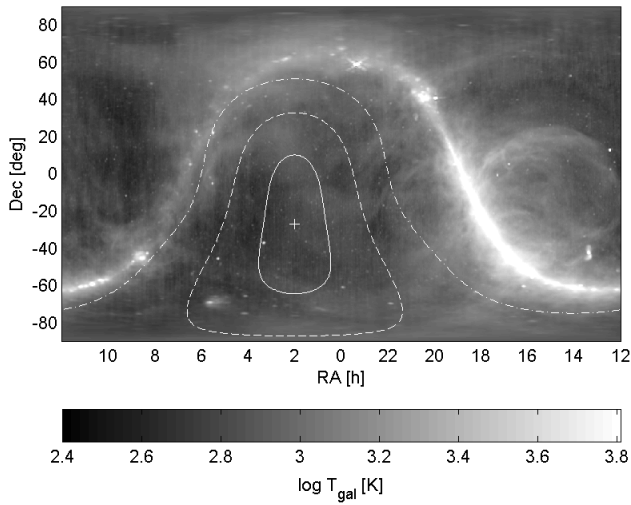}

%\center
%\includegraphics{f7.eps}

\caption[Antenna beam centered at RA=2h projected onto the sky.]{ \label{f_edges_beam} The beam pattern projected onto the sky map of \citet{1982A&AS...47....1H} scaled to 150~MHz using a spectral index of 2.5. The contours are at 90\%, 50\% and 10\% power levels of a dipole over a ground plane. The beam is centered at a right ascension of 2~h.}
\end{figure}

\subsection{Variations in Spectral Index and Temperature}

The spectral index and intensity of the non-thermal contributions to the low-frequency spectrum have been shown to vary across the sky. Drift scan observations with EDGES can measure these variations in $T_{gal}$ (albeit convolved with the large antenna beam) as a function of local apparent sidereal time (LST). In addition, the intensity of the measured spectrum should vary with the polarization direction of the dipole due to differences in the shape of the antenna beam. Sampling both the north-south and east-west polarizations allows additional tests for consistency with expectations from prior measurements.

The radio environment at the Mileura Station is very quiet. In the 100-200 MHz band there are only occasional
very weak signals from distant FM stations via troposcatter, some weak satellite beacons at 150 MHz and occasional
strong signals from the low earth orbit (LEO) satellites in the 137-138 MHz band. The troposcatter comes from the horizon and was not detectable
by the EDGES antenna because the horizon response is more than about 30 dB below the zenith response. The satellite beacons have narrow band CW
signals and their frequencies were excluded from the analysis. The LEO satellites were more of a problem because the signal strengths for
some passes were strong enough to result in some saturation of the 8-bit analog to digital converter. The only recourse was to excise those
time periods. In addition the frequency band from 137-138 was excluded from the analysis.

Four configurations of the system were used during the calibrated observing runs at Mileura Station that provided relevant measurements. Table~\ref{t_edges_cal_runs} lists these configurations.  The observations for each night span about 0~h to 10~h LST. Some data excision has been applied to remove large RFI transient signals. The results of fitting the model for $T_{sky}$ to the high quality, RFI-free data from these runs are shown in Figure~\ref{f_edges_gal_results}.

\begin{deluxetable}{cccl}
    \tablecaption{Sensitivity of spectral index and $T_{150}$ to corrections}
    \tablewidth{0pt}
    \tablehead{
        \colhead{Parameter} &
        \colhead{$\Delta param$} &
        \colhead{$\Delta \beta$} &
        \colhead{$\Delta T_{150}~K$} \\
        }
    \startdata
    loss& 0.1 dB &-0.07& -1 \\
    ambient&+10 C&+0.10& +2 \\
    LNA noise&-40 K& $\approx 0$ & +1 \\
    refl.coeff.&-0.1&+0.09&-7 \\
    \enddata
    \label{t_edges_parm_sens}
%    \tablecomments{ \label{t_edges_parm_sens}
%    }
\end{deluxetable}

The spectral index is typically of order $\beta=2.5$, and is consistent with the measurements of the 1960s and 1970s for this frequency range.  The temperature of the non-CMB contribution at 150~MHz varies with LST, as expected, between approximately 240~K and 300~K over the observed range of LSTs.  The temperature measurements are consistent to within less than 5 K for the north-south polarization measurements with overlapping LSTs, and the temperatures calculated for the east-polarization are about 5 K lower than the equivalent north-south measurements, which is anticipated from extrapolation of all-sky measurements at higher frequencies. The large field of view of the EDGES dipole antenna beam convolved with the sky produces very smooth variations with LST and eliminates much of the structure in both $\beta$ and $T_{150}$ that would be observed with more localized observations.
\begin{deluxetable*}{cccccc}
    \tablecaption{Calibrated Observing Runs}
    \tablewidth{0pt}
    \tablehead{
        \colhead{Date} &
        \colhead{LST} &
        \colhead{Polarization} &
        \colhead{Cable Length} &
        \colhead{Ground Screen} &
        \colhead{Cycle} \\
        \colhead{} &
        \colhead{[h]} &
        \colhead{} &
        \colhead{[feet]} &
        \colhead{} &
        \colhead{[s]}}
    \startdata
    30 Nov 2006 & 0 to 8 & NS & 50 & normal & 210 \\
    01 Dec 2006 & 0 to 8 & NS & 50 & extended & 210 \\
    07 Dec 2006 & 0 to 8 & EW & 100 & normal & 25
    \enddata
    \tablecomments{ \label{t_edges_cal_runs} List of the three calibrated
    observing runs, indicating the configuration of the system during
    each run and the approximate LST of the start and stop times.  The
    polarizations are either approximately north-south (NS) or east-west (EW),
    as described at the beginning of the chapter.
    The cycle times in the last column are the total amount of time spent in
    one complete cycle of the three-position switch.  For the 210~s cycles, the
    division of time between the antenna,
    the ambient load, and the calibrator load is [100, 100, 10]~s, respectively, while for the 25~s cycles,
    it is [10, 10, 5]~s, respectively.}
\end{deluxetable*}

\begin{deluxetable*}{cccccc}
    \tablecaption{Spectral index results}
    \tablewidth{0pt}
    \tablehead{
        \colhead{Reference} &
        \colhead{date} &
        \colhead{Frequency MHz} &
        \colhead{Spectral Index} &
        \colhead{comments}}
    \startdata
    Turtle & 1962 & 100 & $2.5\pm{0.1}$ & \\
           &      & 176 & $2.9\pm{0.1}$ & \\
    Andrew & 1966 & 10-178 & $2.43\pm{0.03}$ & \\
    Purton & 1966 & 10-300 & $2.51\pm{0.05}$ &  \\
    Bridle & 1967 & 81.5  & $2.38\pm{0.03}$ & region of anticenter \\
           &      & 81.5  & $2.46\pm{0.04}$ & region of inner arm \\
    Webster& 1974 & 408-610 & $2.80\pm{0.05}$ & index increases above 400 MHz \\
    Sironi & 1974 & 81.5-408 & $2.41\pm{0.04}$ & average over region out of plane \\
           &      & 151.5-408& $2.49\pm{0.04}$ & \\
    This paper & 2008 & 100-200 & $2.5\pm{0.1}$ & average over region in Figure~\ref{f_edges_beam} \\
            &       & 150-408 & $2.52\pm{0.04}$ & \\
    \enddata
    \label{t_edges_results}
%    \tablecomments{ \label{t_edges_results} Previous results along
%    with the results from this paper.}
\end{deluxetable*}

The most notable irregularity in the measurements is the difference in the variation of the derived spectral index with LST made with and without the extended ground screen.  While the change in estimated loss from the model calculations results in about the same average spectral index around 1 to 3 hours LST the difference in the spectral index derived from data with different ground screen sizes is more significant at other ranges of LST. We take these differences as another indication of the level of uncertainty due to sources of systematic error and note that all the spectral index measurements fall within a range of about 2.4 to 2.6. The most likely explanation for the increase of the spectral index around 8 hours is due to some change in the beam pattern. Figure~\ref{f_edges_beam} shows the beam at 2 hours. In this range the antenna temperature is a minimum and is least sensitive to changes in the beam pattern.

Our final value for the spectral index of the background from
100 to 200 MHz is
\begin{center}
$\beta_{100-200}=2.5\pm{0.1}$.
\end{center}

We can combine our value of minimum non-CMB sky noise at $\nu=150$~MHZ of $(240-2.75)\pm{10}$~K
at $\alpha=2.5$~h with the convolution of the antenna beam with the all sky map (minus the CMB) at $\nu=408$~MHz of
Haslam et al. \citep{1982A&AS...47....1H} to derive a spectral index value of
\begin{center}
$\beta_{150-408}=2.52\pm{0.04}$.
\end{center}

\section{DISCUSSION}

The measurement of the spectral index of the background with a broadband system requires a number of corrections, but
we have shown that it is possible to achieve an accurate result competitive with other radio astronomy techniques. As more broadband systems are built, and the systematics are better understood, there is potential for improvements in the accuracy through better modeling and better antenna design for a wider bandwidth of low reflection coefficient.

Based on the trial measurements with the EDGES system presented in this paper, we are confident that the average spectral index at high Galactic latitudes from 100 to 200~MHz lies between $2.4 < \beta < 2.6$. It should be noted that this is significantly below the the nominal value of $\beta=2.7$ often assumed in this range of frequencies, and a little less than the high Galactic latitude value of $\beta\approx2.6$ that can be extracted from the analysis of \citet{2008arXiv0802.1525D}. It is consistent with many of the earlier findings summarized in Table~\ref{t_edges_results}.

Finally, since much of this analysis was motivated by the needs of the redshifted 21 cm experiments, we would also like to bring attention to the utility of including even one simple, well understood antenna in the large, active dipole arrays being developed.  In addition to being used for absolute measurements of the background, one well calibrated antenna can be used for the calibration of aperture arrays with complex beampatterns using the method described by \citet{1958AuJPh..11...70L}. Such a configuration is sufficient to calibrate the other elements of the array through the redundancy in the correlations on baselines to the single calibrated element and baselines between uncalibrated elements, and can accomplished using only unresolved sources whose flux density need not be known.

\acknowledgments

This work was supported by the Massachusetts Institute of Technology, School of Science, and by the NSF through grant AST-0457585.  JDB is supported by NASA through Hubble Fellowship grant HF-01205.01-A awarded by the Space Telescope Science Institute, which is operated by the Association of Universities for Research in Astronomy, Inc., for NASA, under contract NAS 5-26555.

%\clearpage
%\bibliography{apjmnemonic,../references}
%\bibliography{./references}
%\bibliographystyle{apj}
%\bibliographystyle{apsrev}

\end{document}